\documentclass[aps,pra,twocolumn,showpacs,preprintnumbers,amsmath,amssymb]{revtex4-1}
\usepackage{graphicx} 
\usepackage[usenames]{color}

\begin{document}
\title{Progress towards quantum simulating the classical $O(2)$ model}
\author{Haiyuan Zou$^1$}
\author{Yuzhi Liu$^{2}$}
\author{Chen-Yen Lai$^3$}
\author{J. Unmuth-Yockey$^1$}
\author{Li-Ping Yang$^4$}
\author{A. Bazavov$^{1,3}$}
\author{Z. Y. Xie$^5$}
\author{T. Xiang$^5$}
\author{S. Chandrasekharan$^6$}
\author{S.-W. Tsai$^3$}
\author{Y. Meurice$^1$}
\affiliation{$^1$ Department of Physics and Astronomy, The University of Iowa, Iowa City, Iowa 52242, USA }
\affiliation{$^2$ Department of Physics, University of Colorado, Boulder, Colorado 80309, USA}
\affiliation{$^3$ Department of Physics and Astronomy, University of California, Riverside, CA 92521, USA}
\affiliation{$^4$ Department of Physics,Chongqing University, Chongqing 400044, China}
\affiliation{$^5$ Institute of Physics, Chinese Academy of Sciences, P.O. Box 603, Beijing 100190, China}
\affiliation{$^6$ Department of Physics, Duke University, Durham, NC 27708, USA}
 
\def\lt{\lambda ^t}
\def\note{note}
\def\beq{\begin{equation}}
\def\enq{\end{equation}}

\date{\today}
\begin{abstract}
We connect explicitly the classical $O(2)$ model in 1+1 dimensions, a model sharing important features with $U(1)$ lattice gauge theory, to physical models potentially
implementable on optical lattices and evolving at physical time. 
Using the tensor renormalization group formulation, we take the time continuum limit and check that  finite dimensional projections used in recent proposals for quantum simulators  provide controllable approximations of the original model.
We propose two-species Bose-Hubbard models corresponding to these finite dimensional projections at strong coupling and discuss their possible implementations on optical lattices using a $^{87}$Rb and $^{41}$K Bose-Bose mixture.  
\end{abstract}

\maketitle
\section{Introduction}
\label{sec:intro}
Recently, there has been a lot of interest in the possibility of building quantum simulators for lattice gauge theory (LGT) using optical lattices  \cite{PhysRevLett.110.125303, PhysRevLett.110.125304, Tagliacozzo2013160,Liu:2012dz,Wiese:2013uua}. The goal is to engineer many-body systems with cold atoms that can be built experimentally and that approximately evolve  according to some  given quantum LGT Hamiltonian.  Achieving this goal would allow us to go beyond what can be done with classical computing, namely overcoming the sign problem of Quantum Chromodynamics (QCD) with a chemical potential, establishing its phase diagram and studying its real time evolution. Introducing a chemical potential in QCD is necessary to describe physical situations where a nonzero quark density is needed such as the early universe or heavy ion collisions. 
Building a quantum simulator for QCD requires that we first systematically establish the viability of the approach by building up on  simple models sharing some of the basic features of lattice QCD. 

In the context of condensed matter, a proof of principle that quantum simulating is possible has been given in the case of  the Bose-Hubbard model. For this simple model, a remarkable level of {\it quantitative} agreement \cite{trot}  has been reached between state of the art quantum Monte Carlo 
calculations and their experimental optical lattice implementations. 
It would be very desirable to provide a similar proof of principle in the context of LGT. 

In this article, we propose an optical lattice setup and accurate numerical methods to relate it to a simple model that 
shares some important features (discrete imaginary time, relativistic space-time symmetry, compact gauge variables and a complex action) 
with interesting LGT models,  namely
the classical $O(2)$ in 1+1 dimensions with a chemical potential.  This model is described in section \ref{sec:model}. The goal of the article is to discuss the optical lattice implementation of one of the building block of the Hamiltonian formulation of gauge theory, namely the ``quantum rotors" that are described in more detail below, rather than discussing more specific aspects such as the implementation of Gauss's law for LGT models involving these building blocks.

The connection between the classical $O(2)$ in 1+1 dimensions and physical systems on optical lattices requires three steps. 
First, we introduce new computational methods based on the tensor renormalization group (TRG) method  
\cite{PhysRevB.86.045139,PhysRevD.88.056005,PhysRevE.89.013308,signtrg} to take the time continuum limit (step 1, section \ref{sec:model}) and to calculate the 
effects of finite dimensional truncations necessary for a physical implementation (step 2, section \ref{sec:num}). We then construct a two species Bose-Hubbard model which at second order 
in degenerate perturbation theory can be matched with the finite dimensional truncations and we propose an experimental implementation using a $^{87}$Rb and $^{41}$K Bose-Bose mixture (step 3, section \ref{sec:optical}). The $O(2)$ model is very well understood using classical computing \cite{PhysRevLett.87.160601,Banerjee:2010kc,PhysRevB.86.045139,PhysRevD.88.056005,PhysRevE.89.013308,signtrg}
and our goal is not to learn more about this model from quantum simulations but rather to demonstrate that a quantitative correspondence is possible. 

One should be aware of the fact that in contrast to the quantum Monte Carlo treatment of condensed matter models where space and  time are completely independent entities,  the state of the art calculations in LGT are performed using the Lagrangian formalism at discrete imaginary time 
where  space and time are completely interchangeable. In LGT, the continuum limit is usually taken in a way that preserves this relativistic symmetry between space and time. The Hamiltonian representation provides the functional forms used to fit correlation functions and a slightly better resolution in the time direction is sometimes used, however, the time continuum limit is not  taken independently. Explicit Hilbert space representations of the physical states and of their matrix elements are mostly absent from today's lattice QCD calculations. 
In our construction, the first step will be to take the time continuum limit using the Lagrangian formulation. Note that
Lorentz symmetry can emerge near criticality in the Hamiltonian formulation \cite{PhysRevB.75.134302} and that the classical $O(2)$ model is often used as an effective theory for the Bose-Hubbard model \cite{PhysRevB.40.546}.

It is important to understand the similarity between the infinite dimensional Hilbert spaces of the $O(2)$ model and $U(1)$ LGT in the Hamiltonian formulation. In the mid seventies, LGTs were developed in the Hamiltonian formalism \cite{PhysRevD.11.395,PhysRevD.13.1043,PhysRevD.17.2637,RevModPhys.51.659} using local gauge variables that live on bonds connecting neighboring sites. For continuous and compact symmetry groups, these gauge links are operators  that live on an infinite Hilbert space and in the appropriate basis look like classical group elements. In $U(1)$ LGT \cite{PhysRevD.13.1043}, gauge links are phases $\mathrm{e}^{i\theta}$, which when considered as operators, live in an infinite dimensional Hilbert space spanned by the eigenstates $|n\rangle$ of the ``angular momentum" operator $L=-i\partial/\partial \theta$ with all positive {\it and} negative integer eigenvalues $n$. The same ``quantum rotors" appear in the Hamiltonian formulation of the $O(2)$ model \cite{PhysRevD.17.2637,RevModPhys.51.659}.

For realistic implementations with cold atoms, 
it is necessary to consider Hamiltonians where gauge links are quantum operators that live in a finite rather than infinite Hilbert space \cite{Orland:1989st,Chandrasekharan:1996ih}. In the $U(1)$ example this would mean the eigenvalues of $L$ only take a finite range of values. For this to occur naturally one restricts the Hilbert space to be in a spin-$s$ representation, i.e., $n = -s,-(s-1),..0,..(s-1),s$. Finite dimensional projections and 
quantum link variables have played an important role in recent proposals to simulate dynamical gauge fields \cite{PhysRevLett.110.125303, PhysRevLett.110.125304, Tagliacozzo2013160,cqed,PhysRevLett.109.175302}. 

The common features of the $O(2)$ model considered here, and the $U(1)$ gauge model can be understood by comparing the TRG formulations of the two models \cite{PhysRevD.88.056005}. In both cases, the Fourier expansion of $\exp{(\beta \cos (\theta))}$ is used which leads to the labeling of states by (positive and negative) integers. However, the quantum numbers are associated to plaquettes 
in the gauge case rather than links in the spin case. The physics of the models is also quite different. For instance, in 2+1 dimensions, the 
$O(2)$ spin model has a second order phase transition while the $U(1)$ gauge model has none. 
\section{The model and its time continuum limit}
\label{sec:model}
The simplest model involving the quantum rotors described above is the $O(2)$ model in 1+1 dimensions. 
Its partition function reads: 
\beq
    Z = \int{\prod_{(x,t)}{\frac{d\theta_{(x,t)}}{2\pi}} {\rm e}^{-S}}\ ,
\label{eq:bessel}
\enq
with action 
\begin{eqnarray}
S=&-&  \beta_\tau \sum\limits_{(x,t)} \cos(\theta_{(x,t+1)} - \theta_{(x,t)}-i\mu)\cr&-&\beta_s \sum\limits_{(x,t)} \cos(\theta_{(x+1,t)} - \theta_{(x,t)}).
\end{eqnarray}
The meaning of the chemical potential $\mu$ \cite{Hasenfratz:1983ba} appears clearly in the limit where $\beta_s$ is zero and we have decoupled quantum rotors with a discrete spectrum labeled by $n_x$ at each site $x$ (see Eq. (\ref{eq:tensor})). Using these labels, the chemical potential generates a contribution $-\mu n_x$ to the energy at each site.  
The sites of the rectangular $N_s\times N_\tau$ lattice are labeled as $(x,t)$ and we assume periodic boundary conditions in space and time. 

When $\beta_\tau \gg \beta_s$ we obtain the time continuum limit \cite{PhysRevD.17.2637,RevModPhys.51.659,Polyakov:1987ez} with an Hamiltonian connecting quantum rotors  on a lattice with $\beta_s$ acting as the coupling between the spatial sites. In the $\otimes _x|n_x\rangle$ basis, it reads: 
\begin{equation}
\hat{H}=\frac{\tilde{U}}{2}\sum_x \hat{L}_x^2-\tilde{\mu}\sum_x \hat{L}_x-\tilde{J}\sum_{\left<xy\right>}\cos(\hat{\theta}_x-\hat{\theta}_y) \ ,
\label{eq:rotor}
\end{equation}
with $\tilde{U}=1/(\beta_\tau a)$, $\tilde{\mu}=\mu/a$ and $\tilde{J}=\beta_s/a$, the sum extending over sites $x$ and nearest neighbors $\left<xy\right>$ of the space lattice
and $a$ is a lattice spacing.

The commutation relations $[L,{\rm e}^{\pm i\hat{\theta}}]=\pm {\rm e}^{\pm i\hat{\theta}}$ show that ${\rm e}^{\pm i\hat{\theta}}$ act like creation and annihilation operators. However, there is no eigenstate of $L$ annihilated by ${\rm e}^{- i\hat{\theta}}$. 
At large $\mu$, there is an effective truncation \cite{M.Fisher.2004,sachdev2011} which makes the eigenstates with negative eigenvalues irrelevant. For small value of $\mu$, we will consider 
the quantum link inspired  truncation where the original operator algebra is approximated by a spin-$s$ 
representation with $|n|\leq s$. 

Remembering the role played by the differential operator $L=-i\partial/\partial \theta$ in the construction of the spherical harmonics, we  replace  $L$ by  $L^{3}$, the third component of the angular momentum in the $SU(2)$ Lie algebra. Pursuing the analogy, we replace 
${\rm e}^{\pm i\hat{\theta}}$ by an operator proportional to the raising and lowering operators $ L^{\pm}$ in the spin-$s$ representation. 
In the case of spin-1, a comparison of the matrix elements shows that the correspondence between the two representations can be accomplished by properly choosing the constant of proportionality. 

\section{Numerical calculation of the phase diagram}
\label{sec:num}

We now discuss the phase diagram,  the finite spin projection and the time continuum limit by using the TRG method. 
Following the procedure described in Refs. \cite{PhysRevD.88.056005,signtrg,PhysRevE.89.013308}, we can write 
\begin{equation}
            Z = {\rm Tr} \prod_{(x,t)}T^{(x,t)}_{n_x n_x'n_tn_t'} \ ,
            \label{eq:Z}
\end{equation}
with the local tensor expressed in terms of the modified Bessel functions:
\begin{eqnarray}
\label{eq:tensor}
            T^{(x,t)}_{n_{x}n_{x'}n_{t}n_{t'}} &=&  \sqrt{I_{n_{t}}(\beta_\tau)I_{n_{t'}}(\beta_\tau)\exp(\mu(n_t+n_t'))}\nonumber \\
            & \ &\sqrt{I_{n_{x}}(\beta_s)I_{n_{x'}}(\beta_s)} \delta_{n_{x}+n_{t},n_{x'}+n_{t'}} \ .
        \end{eqnarray}
The indices $n_x, \ n_x',\ n_t$ and  $n_t'$ label the four links coming out of $(x,t)$ in the $x$ and $t$ direction and the trace {\rm Tr} refers to the sum over all these link indices. 
A transfer matrix $\mathbb{T}$ can be constructed by taking the spatial traces in a time slice:  \begin{eqnarray}
&\ &\mathbb{T}_{(n_1,n_2,\dots n_{N_s})(n_1',n_2'\dots n_{N_s}')}=\cr &\ &\cr&\ &\sum_{n_{x1}n_{x2}\dots n_{N_s}} T^{(1,t)}_{n_{xN_s}n_{x1}n_1n_1'}T^{(2,t)}_{n_{x1}n_{x2}n_2n_2'\dots }\cr &\ &\dots  T^{(N_s,t)}_{n_{x(N_s-1)}n_{xN_s}n_{N_s}n_{N_s}' }\ .
\end{eqnarray}
The indices $(n_1,n_2,\dots n_{N_s})$ represent the past and $(n_1',n_2'\dots n_{N_s}')$ the future. 

In view of the rapid decay of the $I_n(\beta)$ when $|n|$ increases at fixed $\beta$, good approximations can be obtained 
by replacing the infinite sums by sums restricted to $-n_{max}$ to $n_{max}$. We denote the number of states $D_{st}=2n_{max}+1$.  With this truncation the transfer matrix is a $D_{st}^{N_s}\times D_{st}^{N_s}$ matrix. It is possible to coarse grain the transfer matrix efficiently by using a higher order singular value decomposition (HOTRG) described in \cite{PhysRevB.86.045139}. This procedure then reduces the two site transfer matrix to a $D_{st} \times D_{st}$ matrix and thus accomplishes the blocking from two sites to a single site. Note that in the spin-1 projection, we keep $D_{st}$ much larger than 3 as we keep blocking. In other words, the spin projection represents a microscopic 
modification of the model, while we need to keep $D_{st}$ as large as possible in order to keep a good macroscopic accuracy. 
The same numerical method is used in all cases, the only difference being the initial tensor. 

An important advantage of the TRG method is that it allows to reach exponentially large volumes. However, it is important to check the results at small volume where sampling methods are feasible and accurate. 
We have used the TRG and the worm algorithm \cite{PhysRevLett.87.160601,Banerjee:2010kc} to calculate the particle number density  \cite{Banerjee:2010kc} 
\begin{equation}
\left<N\right>\equiv1/(N_s\times N_\tau) \partial \ln Z /\partial \mu \ . 
\end{equation}
The partition function $Z$ can be calculated by taking the trace of $\mathbb{T}^{N_{\tau}}$ or by using the methods described in Refs. \cite{PhysRevB.86.045139,PhysRevD.88.056005,PhysRevE.89.013308,signtrg}. 
The numerical values 
on  a $16\times 16$ lattice, 
for values of $\beta_s=\beta_\tau$ and $\mu$ slightly below the 
tips of the  regions with $\left<N\right>=0,\ 1, \dots,\ 4$ of the phase diagram described below are shown in 
Table \ref{tb:comp}.
\begin{table}[h]
\begin{center}
\begin{tabular}{||c|c|c|c||}
\hline
$\beta$ & $\mu$ & $\left<N\right>$ (worm) & $\left<N\right>$(HOTRG) \\
\hline
1.12 & 0.01 & 0.00726(1) & 0.00728(8) \\
0.46 & 1.8  & 0.98929(1)&0.9892(3)\\
0.28&2.85&1.98980(2)&1.989(2)\\
0.2&3.53&2.96646(3)&2.967(1)\\
0.12&4.3&3.96206(4)&3.965(1)\\
\hline
\end{tabular}
\end{center}
\caption{\label{tb:comp} $\left<N\right>$ for the worm algorithm and the HOTRG for $\beta_s=\beta_\tau=\beta$.}
\end{table}

Small discrepancies between the two methods appear typically in the 4th significant digit. 
The errors for the worm algorithm are purely statistical and to the best of our knowledge, there are no systematic errors associated with it. On the other hand, for the TRG method, the limit  $D_{st} \rightarrow \infty$ shows very small variations which 
 will be documented  and analyzed in a separate publication \cite{trgnum} but do not affect the results presented here. 

\begin{figure}[h]
\advance\leftskip -0.7cm
 \includegraphics[width=4.3in]{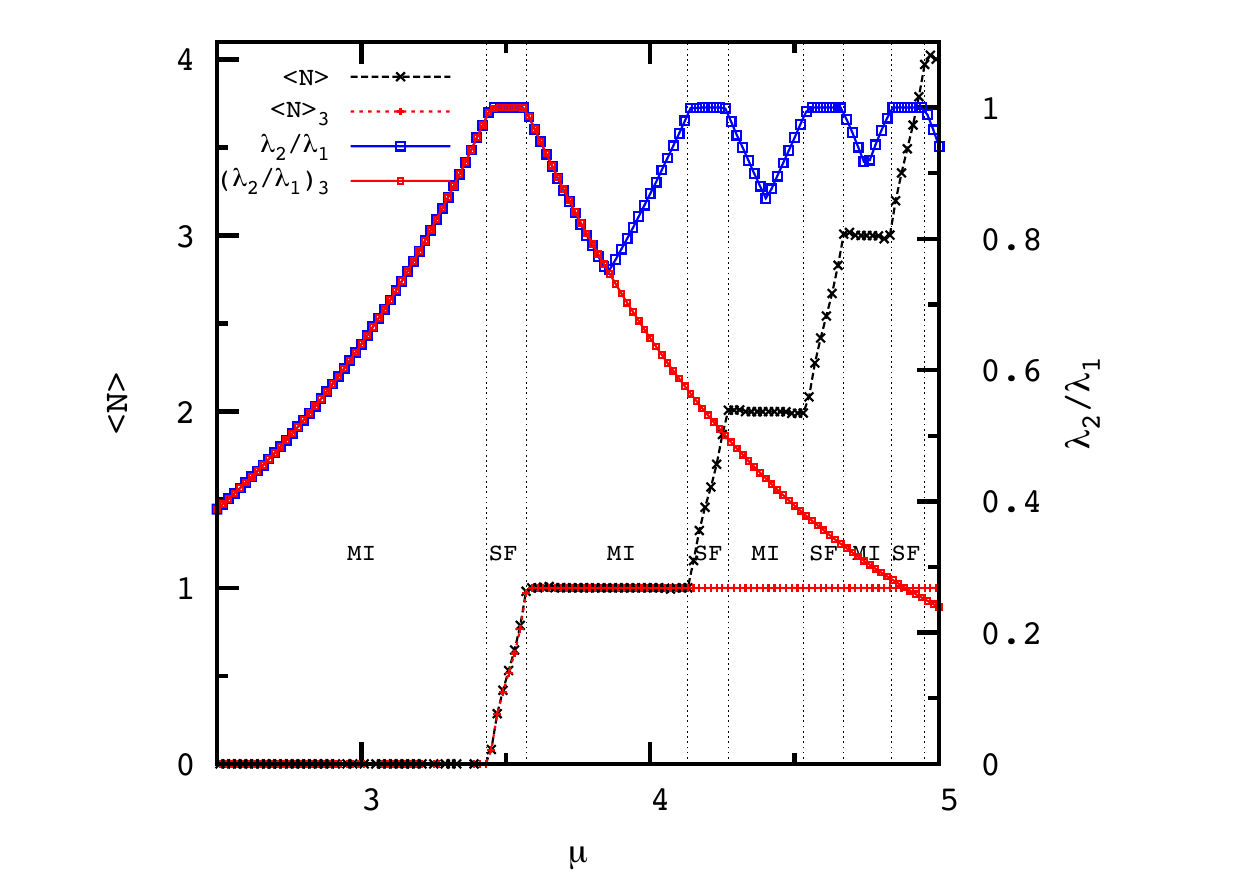}
 \caption{\label{fig:crosspnd} (Color online) Ratio ($\lambda_2/\lambda_1$) of the first two eigenvalues of the transfer matrix and the particle number density $\left<N\right>$ for the  $\beta_\tau=\beta_s= 0.06$ from HOTRG calculation with $D_{st}=15$. The particle number density $\left<N\right>_3$ and the second normalized eigenvalues $(\lambda_2/\lambda_1)_3$ where a lower index 3 denotes the spin-1 projection (3-states) is also shown.}
\end{figure}

By increasing $\mu$ at fixed $\beta$, we go through successive Mott insulating (MI) phases characterized by a fixed integer value of $\left<N\right>$ increasing with $\mu$ and   alternating with superfluid (SF) phases where $\left<N\right>$ interpolates continuously between the successive integers. The phase boundaries are clearly visible from the steps in $\left<N\right>$ as a function of $\mu$ as shown in Fig. \ref{fig:crosspnd}. 
The phase boundaries can also be obtained by looking at the two largest eigenvalues of the transfer matrix. In a given MI phase, one would expect that the largest value of the transfer matrix is unique and corresponds to an eigenstate with fixed integer particle density. On the other hand in the 
the SF phase,  the two largest eigenvalues of the transfer matrix are expected to be degenerate and the corresponding eigenstates to have particle density 
corresponding to the two neighboring MI regions. 
Figure \ref{fig:crosspnd} shows that these expectations are verified quite precisely. 
The system reaches the superfluid (SF) phase when $\lambda_2/\lambda_1=1$ and when $\mu$ is increased further, this ratio stays 1 while there is an increase in the particle number density between two adjacent integers which stand for two different MI phases. 

\begin{figure}[h]
\advance\leftskip -0.7cm
 \includegraphics[width=4.3in]{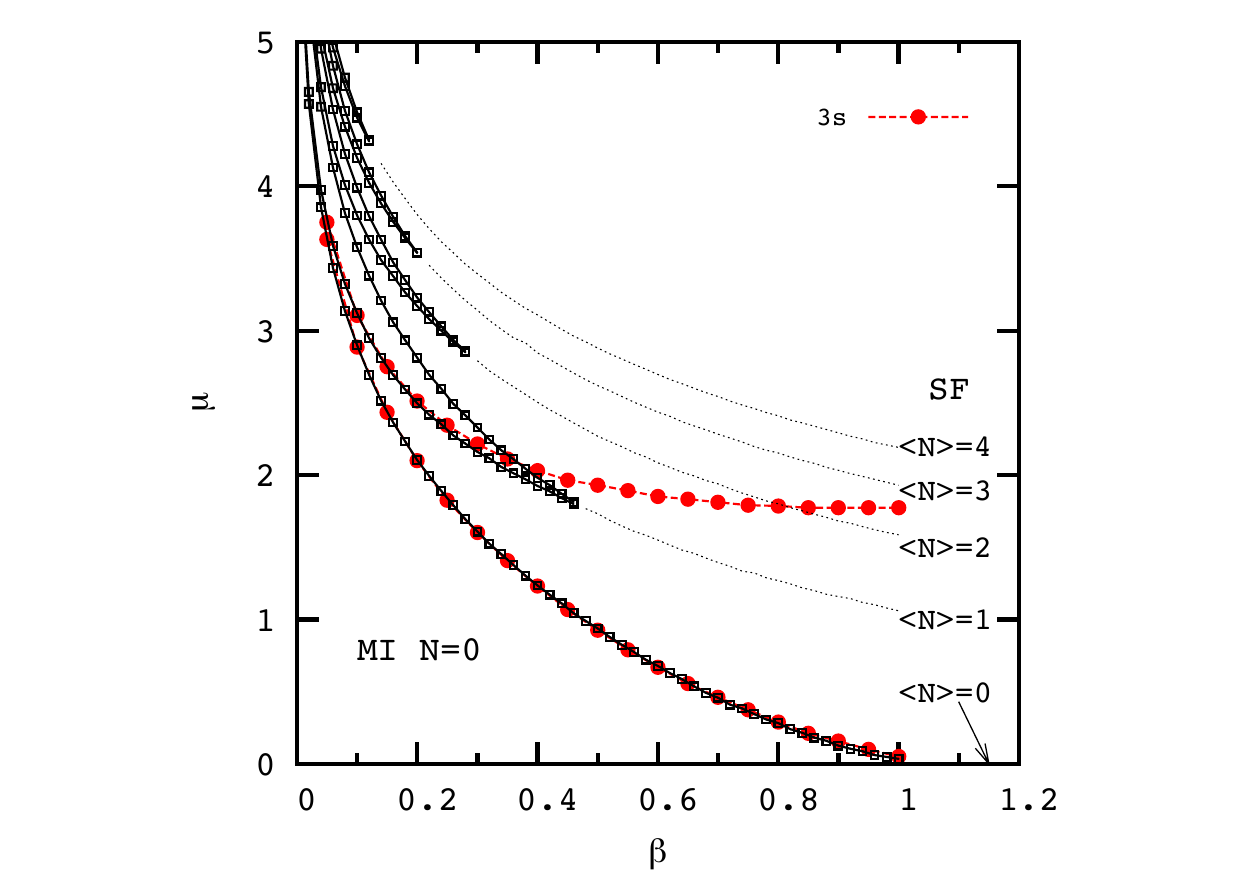}
 \caption{\label{fig:phased2s}  (Color online) The phase diagram in the $\beta$-$\mu$ plane for the isotropic ($\beta_s=\beta_{\tau}=\beta$) case. }
\end{figure}

The alternation between the MI and SF phases in the $\beta$-$\mu$ plane is shown in Fig. \ref{fig:phased2s} . 
The pointy shape of the MI phase region is also observed in other 1+1 dimensional Bose-Hubbard models  \cite{PhysRevB.46.9051,PhysRevB.58.R14741}. The spin-1 projection is also shown in these figures. When $\mu$ is not too large, only small differences with the original, unprojected model are observed. However, when $\mu$ becomes large enough to have $\langle N\rangle> 1$, the truncation prevents such a large occupation and $\langle N\rangle$ saturates to 1 as expected and there is no $\langle N\rangle=2$ MI phase. 
The phase boundary on Fig. \ref{fig:phased2s} between the MI $\langle N\rangle =0$ phase from the SF phase approximately coincides with the line for the model with an infinite number of states. 
Similarly, the spin-2 projection (not shown on the graph) reproduces well the $\langle N\rangle =0$ and 1 boundaries while discrepancies appear for $\langle N\rangle =2$. 

Figure \ref{fig:phased2s} shows that when $\beta$ is small, the boundary between the MI and SF phase appear to be at large values of $\mu$. It is useful to recall that so far we have only considered the phase diagram in the isotropic case $\beta=\beta_s=\beta_\tau$. When $\beta \rightarrow 0$, the interaction along the space links are small, but if $\mu$ is sufficiently large, the interactions along the time links are not small. In the limit where the interactions among the space links are negligible, the problem reduces to a collection of independent one site problems (simple quantum mechanics) as in mean field theory \cite{PhysRevB.40.546}. In this limit, Eq. (\ref{eq:tensor}) shows that the transfer matrix becomes diagonal because $I_n(0)=0$ except for $n=0$ ($I_0(0)=1$) and by the conservation law the same index $n_x$ characterizes the interaction along the time direction.  In other words, there is no quantum number flowing in the space direction and the flow in the time direction at each site is constant. In this limit, the eigenvalues of the transfer matrix are just 
\begin{equation}
\lambda_{(n_1,n_2,\dots n_{N_s})}=
\prod_x I_{n_x}(\beta)e^{{n_x}\mu}\ .
\end{equation}
The largest eigenvalue is then 
\begin{equation}\lambda_{max}=({\rm Max_n}(I_{n}(\beta)e^{{n}\mu}))^{N_s}. \end{equation}
Finding the value of $n$ corresponding to the maximum eigenvalue gives the particle density $\left<N\right>$ in the MI phase. 

\begin{figure}[h]
\advance\leftskip -0.7cm
 \includegraphics[width=4.3in]{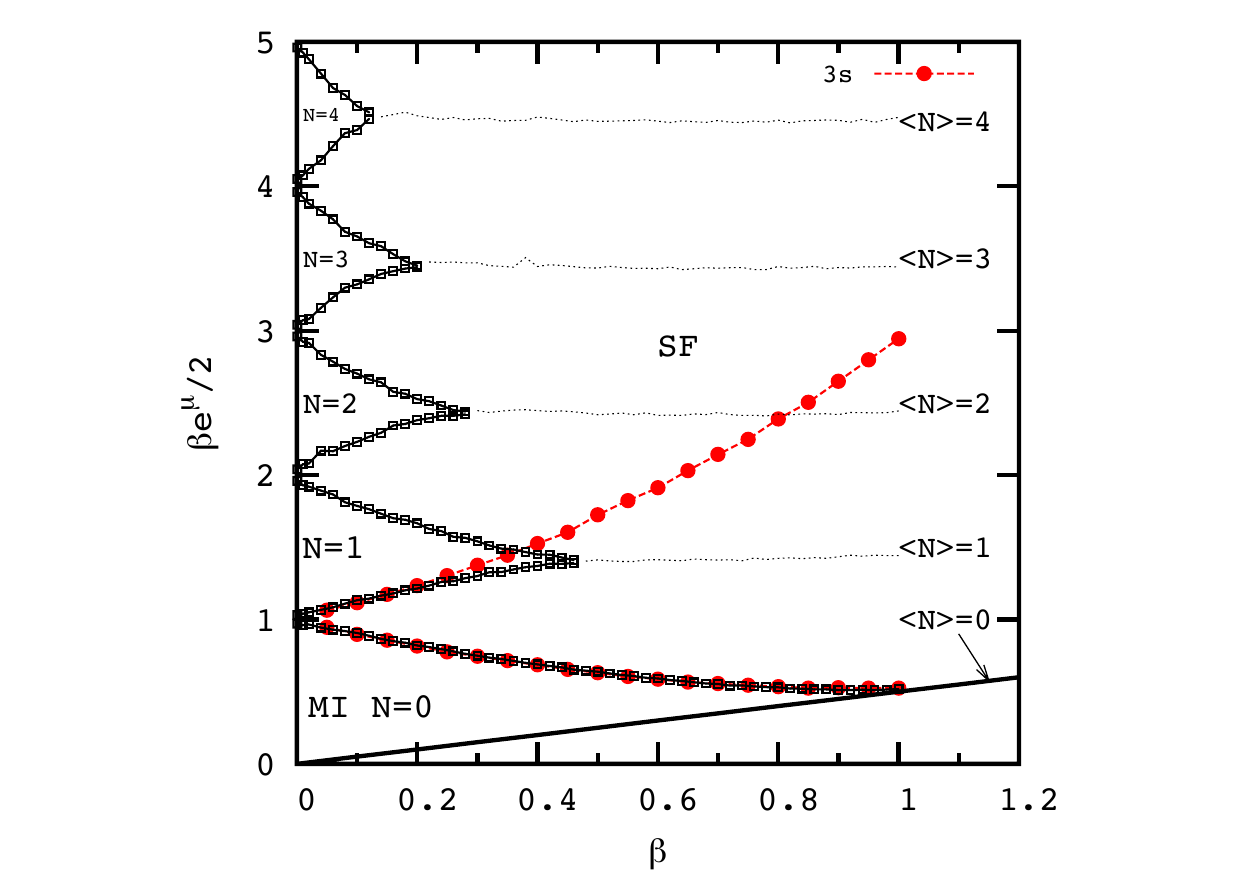}
 \caption{\label{fig:phased2sB}  (Color online) The phase diagram in the $\beta$-$\beta e^{\mu}/2$ plane for the isotropic case. $\left<N\right>=0$ ($\mu=0$) line in the SF phase is the Kosterlitz-Thouless phase in the 1+1 D  $O(2)$ model at $\mu =0$. The lines labeled by ``3s" stand for the phase boundaries of the spin-1 (3-states) system. }
\end{figure}

The maximization of $A_n=I_n(\beta)e^{n\mu}$ can be achieved by considering the ratios $r_n=A_{n+1}/A_{n}$.
Note that we assume $\mu >0$ and given that $I_n(\beta)=I_{-n}(\beta)$, we only need to consider $n\geq0$. 
When $r_{n-1}>1$ and $r_{n}<1$, $A_n$ is a maximum. 
It can be shown in the limit of small and large $\beta$  that $r_n$ decreases when $n$ increases. If this remains true for arbitrary $\beta$ and if $r_n \neq1$, then the problem has a unique solution. 
The interesting case is $r_n=1$ which implies $A_n=A_{n+1}$ and should be at the boundary between two MI phases with particle density $n$ and $n+1$. In the small $\beta$ limit, $I_n(\beta)\simeq \beta^n/(2^nn!)$ and the condition $r_n=1$ implies that 
\begin{equation}\label{eq:deg}\beta e^{\mu}/2 =n+1\end{equation} in that approximation.  The sudden transition in particle density occurs at integer values of  $\beta{\rm e}^\mu /2$. This prediction is confirmed by plotting the phase diagram in the $\beta$-$\beta e^{\mu}/2$ plane as shown in Fig. \ref{fig:phased2sB}. We see that by changing the vertical coordinate to $\mu\rightarrow\beta e^{\mu}/2$, the shape of the phase diagram of the isotropic system looks like the cuspy shapes found for the Bose-Hubbard model in one spatial dimension \cite{PhysRevB.46.9051,PhysRevB.58.R14741}. Keeping in mind that we are working in the limit of small $\beta$, Eq. (\ref{eq:deg}) implies that the phase boundaries of the SF phase between the $n$ and $n+1$ MI phases diverge like $\ln (2(n+1)/\beta)$ when $\beta\rightarrow 0$ which is consistent with Fig. \ref{fig:phased2s}.

\begin{figure}[h]
\advance\leftskip -0.7cm
 \includegraphics[width=4.3in]{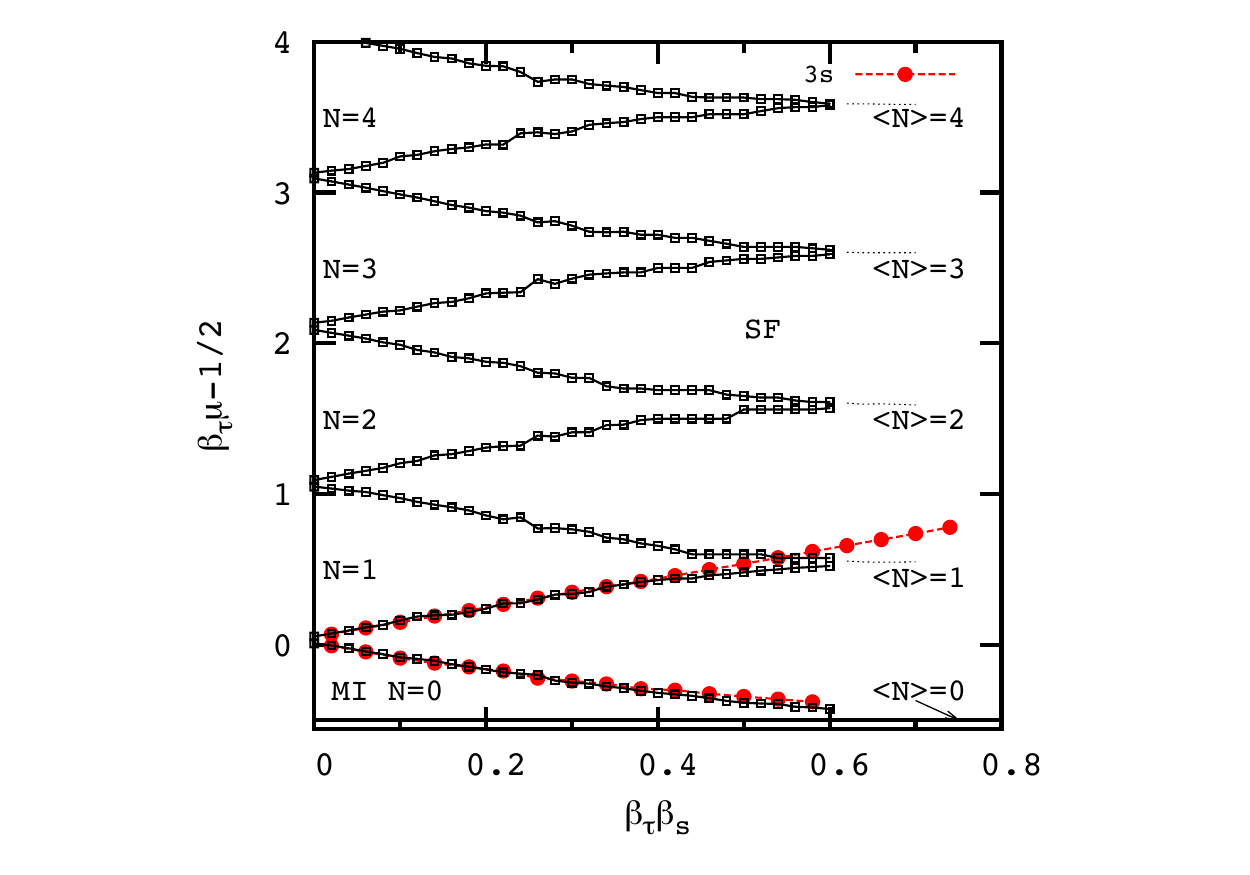}
 \caption{\label{fig:phased2a}  (Color online) Phase diagram for the 1+1 D $O(2)$ model at $\beta_\tau=10$ in $\beta_e$-$\mu_e$ plane is shown. The lines labeled by ``3s" stand for the phase boundaries of the spin-1 (3-state) system.}
\end{figure}

We now depart from the isotropic $\beta_\tau=\beta_s$ situation and consider the case $\beta_\tau\gg \beta_s$
corresponding to the time continuum limit. If we neglect $\beta_s$ we obtain the one site approximation described above. 
The particle density can be obtained from the ratio analysis in the large $\beta_\tau$ limit. Using $I_{n+1}(\beta_\tau)/I_{n}(\beta_\tau)\simeq 1-((n+1/2)/\beta_\tau$ in this limit, we find that the degeneracy occurs for integer values of $\mu\beta_\tau-1/2$.
Defining the effective chemical potential $\mu_e=\mu\beta_\tau-1/2$ and effective coupling $\beta_e=\beta_s\beta_\tau$, we find that the same MI-SF pattern appears in the $\beta_e$-$\mu_e$ plane (Fig. \ref{fig:phased2a}). 

Having computed the phase diagram in the $\beta_\tau=\beta_s$ and $\beta_\tau\gg \beta_s$ cases, we learned that they have very similar shapes in suitable systems of coordinates. From this we expect that they can be smoothly deformed into 
each others and that nothing special happens in the intermediate situations.

\section{Optical lattice implementation}
\label{sec:optical}

In order to incorporate the positive and negative eigenvalues of $L$, we will consider a two-species Bose-Hubbard Hamiltonian on a lattice:
\begin{equation}\label{2bh}
  \begin{split}
\mathcal{H}&=-\sum_{\langle xy\rangle}(t_a a^\dagger_x a_y+t_b b^\dagger_x b_y+h.c.)-\sum_{x, \alpha}(\mu + \Delta_\alpha) n^\alpha_x\\
&+\sum_{x, \alpha}\frac{U_\alpha}{2}n^\alpha_x(n^\alpha_x-1)+W\sum_xn^a_xn^b_x+\sum_{\langle xy\rangle,\alpha}V_\alpha n^\alpha_xn^\alpha_y \,
  \end{split}
\end{equation}
with $\alpha=a,b$ indicating the two different species, 
 $n^a_x=a^\dagger_xa_x$ and $n^b_x=b^\dagger_xb_x$ the number operators, and $|n^a_x,n^b_x\rangle$ the corresponding on-site basis. 
 This class of models has been studied extensively  
 \cite{Kuklov:2003bra,Kuklov:2004hl,Chen:2010is,Chung:2012cb}. 
It is possible to adjust the chemical potentials in order to set $\langle n_x\rangle = \langle n_x^a + n_x^b\rangle = 2$. 
 In the limit where $U_a=U_b=W$ are very large and positive, 
the on-site Hilbert space can then be restricted to the states satisfying $n_x=2$ at each site. All the other states (with $n_x\ne 2$) belong to high-energy sectors that are separated from this one by energies of order $U$. The three states 
 $|2,0\rangle$, $|1,1\rangle$ and $|0,2\rangle$ 
 correspond to the three states of the spin-1 projection considered above. 
 
 It is useful to visualize the minima of the on-site Hamiltonian obtained in the limit $t\rightarrow 0$. 
 It can be written as a quadratic form and a linear term in $n_a$ and $n_b$. If $U_aU_b>W^2$, there is a unique 
 minimum, $|1,1\rangle$, which corresponds to a miscible phase where the two species need to be present at the same place. Since 
 in the spin-1 approximation, $|1,1\rangle$ corresponds to a rotor with angular momentum zero, this is the correct situation for the $O(2)$ model we try to simulate. 
 On the other hand, if $U_aU_b<W^2$, the extremum is a saddle point. As we will discuss later, the unstable direction coming out of the extremum is limited by the positivity of the occupation number. There are two vacua $|2,0\rangle$ and $|0,2\rangle$, which corresponds to immiscible phases.  The limiting case $U_aU_b=W$ corresponds to our $U_a=U_b=W=U_0$ lowest order approximation. If in addition we have $\mu=(3/2)U_0$ and $\Delta_\alpha=0$, we have a flat direction  along the line $n_x=2$ where we have three states of energy $-2U_0$, 
while the degenerate lines with $n_x=1$ or 3 have energy $-(3/2)U_0$ which is considered much larger in the strong coupling approximation. Small changes in the parameters will break the degeneracy of the ground state but preserve a significant difference between these states and the excited states. Decreasing $W$ lowers the energy of the $|1,1\rangle$  state linearly in the difference with $U_0$. Similarly, increasing $W$ raises the energy of the 
$|1,1\rangle$, the flat direction curves down at both ends, but the positivity of the occupation number prevents to have energy unbounded from below. When species dependent chemical potentials are turned on, the flat direction becomes slanted linearly in the variation of the chemical potential $\Delta_\alpha$. 
The overall shape of the trap will typically create small variations in a space-dependent manner. 
In summary, as long as the variations of the parameters are small compared to $U_0$, the features departing from the degenerate case can be treated as perturbations. 
 
 Going back to the general Hamiltonian (Eq. \ref{2bh}), we write $U_{a(b)} = U \pm \delta$ and 
 assume $U\gg\delta, (U-W), V, t_{\alpha}, \Delta_{\alpha}$ and do degenerate perturbation theory.
Virtual processes exchanging particles between neighboring sites are allowed at second order with contributions proportional to $-t_\alpha t_{\alpha^\prime}/U$. 
The hopping amplitude is tunable and when chosen to be $t_\alpha=\sqrt{V_\alpha U/2}$, the final result is that the effective Hamiltonian up to  second order in degenerate perturbation theory corresponds to the spin-1 projection of the rotor Hamiltonian of Eq. (\ref{eq:rotor}) with 
$\tilde{J}=\sqrt{V_aV_b}$, $\tilde{U}=2(U-W)$, and $\tilde{\mu}=(\Delta_a-V_a)-(\Delta_b-V_b)$. Similarly, by increasing the chemical potentials, it is possible to restrict the Hilbert space to $n^a_x+n^b_x=2s$ which corresponds to a spin-$s$ projection in the $O(2)$ model.

\begin{figure}[!t]
  \begin{center}
    \includegraphics[width=0.5\textwidth]{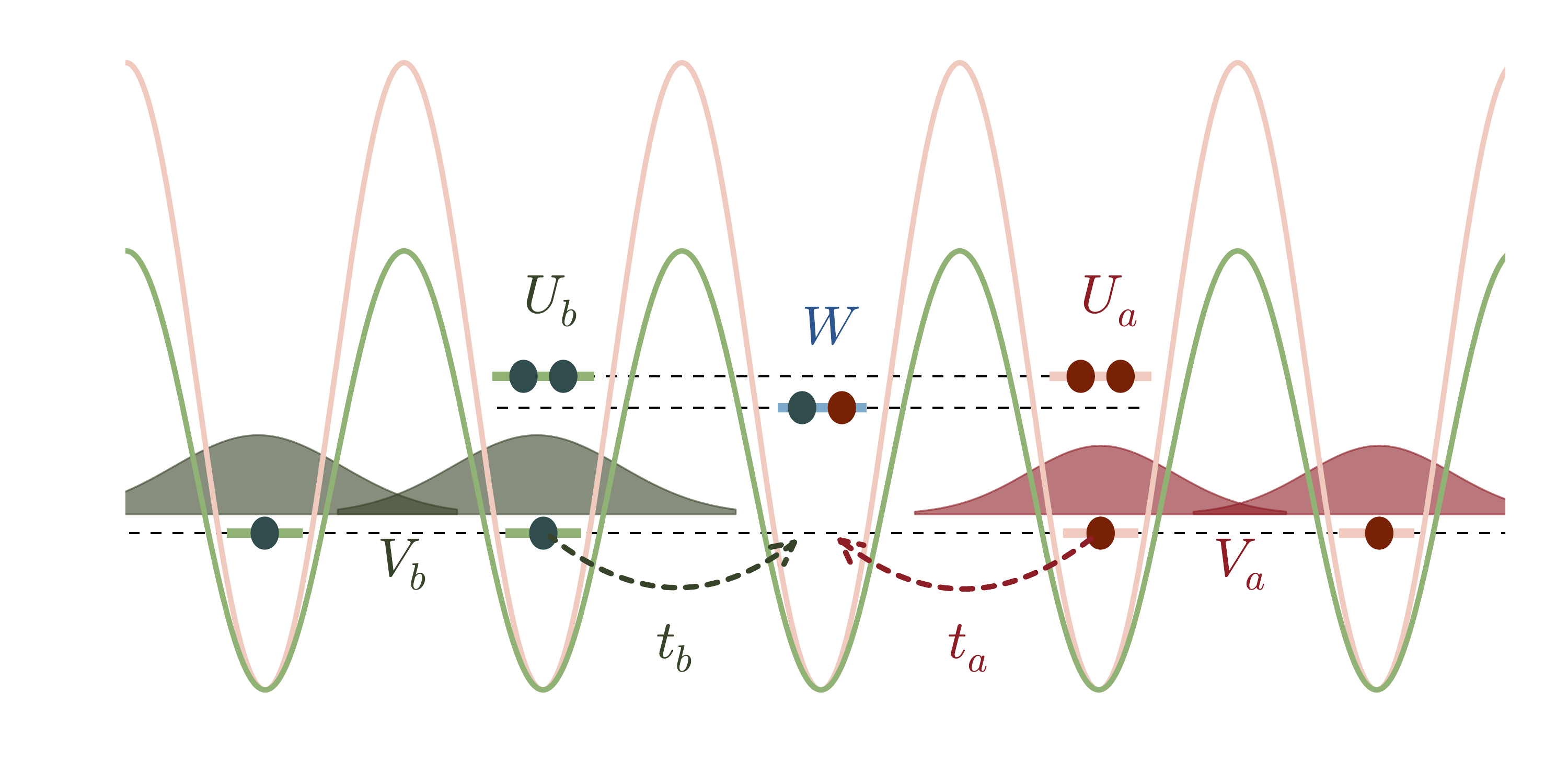}
    \caption{\label{fig:optical} (Color Online) Two-species (green and red) of bosons 
    on species-dependent optical lattices (with the same color). The nearest neighbor interaction is coming from the overlap of Wannier Gaussian wave functions. We assume the difference between intra-species interactions are small $U\gg\delta$.}
  \end{center}
\end{figure}

This two-species Bose-Hubbard model can be realized in a $^{87}$Rb and $^{41}$K Bose-Bose mixture where an interspecies Feshbach resonance is accessible \cite{Catani:2008fp,Thalhammer:2008in}. 
Due to the physical nature of the different atoms, the hopping amplitudes ($t_a, t_b$) are different to begin with,
as well as the intraspecies interactions. 
In addition, species-dependent optical lattices \cite{Lee:2007ip,Anderlini:2007eb,McKay:2010jn,SoltanPanahi:2011ey,Belmechri:2013fe}  are widely used in boson systems, which allows the hopping amplitudes of each individual species to be further tuned to the desired value. 
As mentioned above, the interspecies interaction ($W$) can be controlled by an external magnetic field \cite{Thalhammer:2008in}. 
Finally, the extended repulsion, $V_\alpha$, is present and small when we consider Wannier Gaussian wave functions centered on nearby lattice sites according to previous study \cite{Mazzarella:2006cx}. This is schematically illustrated in Fig. \ref{fig:optical}. 
This may be the most difficult parameter to achieve, but other proposals may be explored, such as by using dipolar bosons \cite{Trefzger:2009cx}, or by pumping bosons to higher Bloch bands \cite{Scarola:2005vq} in order to engineer the nearest neighbor interaction. It is also important to have $U$ significantly larger than the temperature. For the mixture considered here the temperature  and recoil energies are of the order of 100nK and values of $U$ 10-20 times larger can typically be reached \cite{Thalhammer:2008in,PhysRevA.85.023623,RevModPhys.80.885}.

\section{Conclusions}
In summary, we have used new numerical methods to connect the $O(2)$ model in 1+1 dimensions to an optical lattice setup. 
A  first test of the correspondence  would be to check that the optical lattice system reproduces the phase diagram of Fig. \ref{fig:phased2a} 
which corresponds to the time continuum limit $\beta_\tau \gg \beta_s$ of the classical model and where the microscopic parameters can be approximately 
connected to those of the two-species Hubbard model. 

The TRG method presented here allows reliable calculations of the eigenvalues $\lambda_i$ of the transfer matrix. In the time continuum limit, 
we have 
\begin{equation}
\lambda_i/\lambda_1 \propto {\rm e}^{-a(E_i-E_0)} \ ,
\end{equation}
with $E_i$ the corresponding energies and $a\propto 1/\beta_\tau$ the lattice spacing. Recently developed experimental techniques, e. g. momentum resolved Bragg spectroscopy \cite{bragg}, could in principle allow detailed comparisons. 

We have shown that for low enough $\mu$, the effect of the truncation to spin-1 or 2 of the original $O(2)$ model had small effects on the phase boundaries. In the TRG 
formulation, this truncation only affects the initial values of the tensor which can be compared with the initial tensor of other spin models with a finite number of states in 
configuration space (clock and Potts models). Understanding how the symmetries of this initial tensor affects the universality class is under study. 

The $O(2)$ model has an exact conservation law which is made clear by the Kronecker delta in Eq. (\ref{eq:tensor}). 
The states kept in the TRG calculation have a well-defined quantum number associated with this conservation law and it can monitored and  put into histograms \cite{trgnum}. This provides detailed information about the average occupation and its fluctuations. It could give a better insight into the validity of Gutzwiller ansatz or mean field calculations such as the ones discussed in Ref.  \cite{PhysRevB.40.546}, or the validity of the finite spin projection discussed here. 

In LGT calculations, important information regarding the spectrum and matrix elements can be extracted from the 2 and 3 point functions obtained by introducing localized sources in the Lagrangian formulations. Techniques to gather related information from an optical lattice system remain to be developed. Generalizing the work done here for the $O(3)$ model which has 
a physics more similar to lattice QCD seems possible and interesting. 

Acknowledgments. We thank Masanori Hanada, Peter Orland, Lode Pollet, Boris Svistunov, the participants of the Math-Physics and Particle and Nuclear Seminar at the University of Iowa and of SIGN 2014  for stimulating discussions. This research was supported in part  by the Department of Energy
under Award Numbers DOE grant DE-FG02-05ER41368, DE-SC0010114 and DE-FG02-91ER40664, the NSF under grant DMR-1411345 and by the Army Research Office of the Department of Defense under Award Number W911NF-13-1-0119.  Part of the simulations were done at CU Boulder Janus clusters and Y. L. thanks D. Mohler and J. Simone for discussions on code development. 

%

\end{document}